\documentclass{aa}
\usepackage{graphics}
\begin{document}

\thesaurus{11.02.1; 11.02.2; 13.25.2}

\title{The concave X--ray spectrum of the blazar ON\,231:
the signature of intermediate BL Lac objects}

\author{}
\institute{}

\author{G. Tagliaferri\inst{1} \and G. Ghisellini\inst{1} 
\and P. Giommi\inst{2} 
\and L. Chiappetti\inst{3} 
\and L. Maraschi\inst{4}
\and A. Celotti\inst{5} 
\and M. Chiaberge\inst{5} 
\and G. Fossati\inst{6} 
\and E.Massaro\inst{7}
\and M. Maesano\inst{7}
\and F. Montagni\inst{7} 
\and R. Nesci\inst{7}
\and G. Nucciarelli\inst{8} 
\and E. Pian\inst{9}
\and C.M. Raiteri\inst{10}
\and F. Tavecchio\inst{4} 
\and G. Tosti\inst{8}
\and A. Treves\inst{11}
\and M. Villata\inst{10}
\and A. Wolter\inst{4}
}

\offprints{G. Tagliaferri}

\institute{Osservatorio Astronomico di Brera, Via Bianchi 46, I-23807 Merate,
Italy
\and {\it Beppo}SAX Science Data Center, ASI, Via Corcolle, 19, 
          I-00131 Roma, Italy
\and Istituto di Fisica Cosmica G.Occhialini, CNR, Via Bassini 15, I-20133 Milano, Italy
\and Osservatorio Astronomico di Brera, Via Brera, 28, I-20121 Milano, Italy
\and SISSA/ISAS, via Beirut 2-4, 34014 Trieste, Italy
\and CASS, University of California at San Diego, 9500 Gilman Drive, 
     La Jolla, CA 92093-0424, USA
\and Istituto Astronomico, Universit\`a ``La Sapienza'', Via Lancisi 29,
    I-00161 Roma, Italy
\and Osservatorio Astronomico, Universit\`a di Perugia, Via A. Pascoli,
     I-06100 Perugia, Italy 
\and ITeSRE/CNR, Via Gobetti 101, I-40129 Bologna, Italy
\and Osservatorio Astronomico di Torino, Strada Osservatorio 20, I-10025
     Pino Torinese, Italy
\and Istituto di Fisica, Universit\`a dell'Insubria, Via Lucini 3, I-22100 Como, Italy
}

\date{Received date / accepted date}

\titlerunning{The concave X-ray spectrum of ON\,231}
\authorrunning{Tagliaferri et al.\ }

\maketitle

\begin{abstract}

{
ON\,231 was observed with $Beppo$SAX in May and June 1998, 
following an exceptional optical outburst which occured in
April--May. We measured the X--ray spectrum from 0.1 up to 100 keV.
In both occasions the spectrum had a concave shape, with a break 
detected at about 4 and 2.5 keV, respectively.
We interpret the steeper component at energies below the break as 
due to synchrotron emission and the extremely flat component 
at energies above the break as due to inverse Compton emission. 
This is so far the best example in which both the synchrotron
and the Inverse Compton component are detected simultaneously
and with the same instruments in the X--ray spectrum of a blazar.
We detect a decrease of the X--ray flux of about 30\% below the break 
between the first and the second observation, and smaller
variability amplitude between 4 and 10 keV.
During the May observation we also detected a fast variability event
with the flux below 4 keV increasing by about a factor of three in
5 hours. Above 4 keV no variability was detected.
We discuss these results in the framework of synchrotron
self--Compton models.
}
                
\end{abstract}

\section{Introduction}

The continuum emission from Active Galactic Nuclei (AGN) is both highly
luminous and rapidly variable, especially for the blazar class (BL Lac
objects and violently variable quasars). 
Determining the continuum production mechanism is critical for understanding 
the central engine in AGNs, a fundamental goal in extragalactic astrophysics. 
The observed radiation of blazars is dominated by the emission of a jet
whose plasma moves relativistically at small angles to the line of sight
(Blandford \& Rees 1978). Early multiwavelength
studies provided the first strong evidence for bulk relativistic motion, 
later confirmed directly with VLBI observations (Vermeulen \& Cohen 1994).
However, single epoch spectra cannot constrain the models of variability
in relativistic jets (e.g. K\"onigl 1989; Ulrich et al. 1997).

The overall spectral energy distribution (SED) of blazars
shows two broad emission peaks: the lower frequency peak is believed to be 
produced by synchrotron emission, while the
higher frequency peak should be due to the inverse Compton process.
The location of the synchrotron peak is used to define different 
classes of blazars: HBL (High frequency peak blazar,
peaking in the UV or X--ray frequencies) and LBL 
(Low frequency peak blazar, peaking in the IR or optical bands)
(Giommi \& Padovani 1994).

Since blazars emit over the entire electromagnetic spectrum, a key for 
understanding blazar variability is the acquisition of several wide
band spectra in different luminosity states during major flaring episodes.
Coupling spectral and temporal information greatly constrains the jet
physics, since different models predict different variability as a
function of wavelength.
Important progress in this respect has been achieved recently for
some of the brightest and most studied blazars, as PKS\,2155-304 
(Chiappetti et al. 1999; 
Urry et al. 1997), BL\,Lac (Bloom et al. 1997), 3C\,279 (Wehrle et al. 1998),
Mkn\,501 (Pian et al. 1998), Mkn\,421 (Maraschi et al. 1999).
We successfully used the {\it Beppo}SAX satellite to perform observations 
of blazars that were known to be in a high state from observations carried 
out both in other bands (mainly optical and TeV) and in the X--ray band 
itself.
The good {\it Beppo}SAX sensitivity and spectral resolution over a very wide 
X--ray energy range (0.1--200 keV) are ideal to constrain the existing 
models for the X--ray emission.

The BL\,Lac object \object{ON\,231} (\object{W\,Com}, \object{B2\,1219+28}, 
$z=0.102$), which had been observed in the X--ray band by
{\it Einstein} IPC in June 1980 with a 1 keV flux of
$1 \mu$Jy (Worrall \& Wilkes 1990) and by ROSAT PSPC in 
June 1991 with a 1 keV flux of $0.4 \mu$Jy and energy spectral
index $\alpha = 1.2$ (Lamer et al. 1996, Comastri et al. 1997),
had an exceptional optical outburst in April--May 1998,
reaching the most luminous state ever recorded, about 40  mJy
in the R band. The optical broad band spectrum was strongly variable.
In particular, it was very flat at the maximum with a broad band 
energy spectral index of 0.52, while before the flare it was found
to be 1.4; the peak 
frequency moved from near IR to beyond the B band. During
the flare a sudden and large increase of the linear polarisation,
from about 3\% to 10\%, was also observed and it remained high at
least to the end of May, indicating a non-thermal origin of the burst
(Massaro et al. 1999). An optical spectrum of ON\,231 was obtained by
Weistrop et al. (1985) and it shows two weak emission features identified
with the H$_{\alpha}$ (EW ~ 1 or 2 \AA) and O\,III, from which a redishift
estimate of z=0.102 was derived. No more recent spectra, in particular 
during the flare, have been published.
Following the optical flare, we triggered our X--ray observation and
ON\,231 was observed by {\it Beppo}SAX in May, with a second pointing
performed a month later, in June. 
We measured for the first time the hard X--ray spectrum of this 
source above 3 keV and in different brightness states. 
In these occasions simultaneous optical observations were also performed.
Unfortunately the source was already close to the sun and it was impossible
to monitor it long enough to search for correlated variability at 
optical and X--ray frequencies.

In this paper we present and discuss the results of these
$Beppo$SAX observations together with simultaneous optical data.

\section{X-ray Observations}

\subsection{Observations and Data Reduction}

The {\it Beppo}SAX satellite is the result of an international
collaboration between the Italian Space Agency (ASI), the Netherlands
Agency for Aerospace Programs (NIVR) and the Space Science Department
of the European Space Agency (SSD-ESA). It carries on board four Narrow
Field Instruments (NFI) pointing in the same direction and covering a 
very large energy range from 0.1 to 300 keV (Boella et al.\ \cite{BBP97a}).
Two of the four instruments have imaging capability, the Low Energy
Concentrator Spectrometer (LECS), sensitive in the range 0.1--10 keV
(Parmar et al.\ \cite{PMB97}), and the three Medium Energy Concentrator
Spectrometers  (MECS) sensitive in the range 1.3--10 keV (Boella
et al.\  \cite{BCC97b}). The LECS and three MECS detectors are all Gas
Scintillation Proportional Counters and are at the focus of four identical
grazing incidence X--ray telescopes. The other two are passively collimated
instruments: the High Pressure Proportional Counter (HPGSPC), sensitive
in the range 4--120 keV (Manzo et al. 1997) and the Phoswich Detector
Systems (PDS), sensitive in the range 13--300 keV (Frontera et al. 1997). 
For a full description of the {\it Beppo}SAX mission see 
Boella et al. (1997a).

\begin{table*}
\caption{Journal of observations}
\begin{center}
\begin{tabular}{|l|cccccc|}
\hline
date & LECS  & count rate$^{\rm a}$ & MECS & count rate$^{\rm b}$ & 
PDS & count rate$^{\rm c}$ \\
& exposure (s) & $cts \ s^{-1}$ & exposure (s) & $cts \ s^{-1}$ & 
exposure (s) & $cts \ s^{-1}$ \\
\hline
& & & & & &\\
11-12/5/1998 & 21127 & $0.083 \pm 0.002$ & 24859 & $0.075 \pm 0.002$ & 40729 
& $0.24 \pm 0.04$ \\ 
11-12/6/1998 & 17673 & $0.054 \pm 0.002$ & 36661 & $0.052 \pm 0.002$ & 32482 
& $0.12 \pm 0.05$ \\
\hline \hline
& V mag & R$_{\rm c}$ mag & I$_{\rm c}$ mag \\
\hline
11/5/98  & $13.39 \pm 0.04$ & $13.07 \pm 0.02$ & $12.54 \pm 0.03$ & & & \\
12/5/98  & $13.64 \pm 0.03$ & $13.25 \pm 0.02$ & $12.73 \pm 0.02$ & & & \\
11/6/98  & $13.31 \pm 0.03$ & $12.92 \pm 0.03$ & $12.38 \pm 0.03$ & & & \\
13/6/98  &                  & $12.93 \pm 0.03$ &                  & & & \\
\hline
\end{tabular}
\end{center}
$^{\rm a}$ In the band 0.1--10 keV;
$^{\rm b}$ for two MECS units in the band 1.5--10 keV; 
$^{\rm c}$ in the band 12--100 keV.
\end{table*}

The log of the ON\,231 observations is given in Table 1, together with
the exposures and the mean count rates in the various instruments. The data
analysis for the LECS and MECS instruments was based on the linearized,
cleaned event files obtained from the online archive (Giommi \& Fiore 
\cite{GF97}). Light curves and spectra were accumulated with the FTOOLS
package (v. 4.0), using an extraction 
region of 8.5 and 4\,arcmin radius for the LECS and MECS, respectively.
At low energies the LECS has a broader Point Spread Function (PSF) than the 
MECS, while above 2\,keV the PSFs are similar. The adopted regions provide 
more than 90\% of the source counts at all energies both for the LECS and 
MECS. 
The LECS and MECS background is low and rather stable, but not uniformly 
distributed across the detectors. 
For this reason, it is better to evaluate the
background from blank fields, rather than in concentric rings around the 
source region.
Thus, after having checked that the background was not varying during the 
whole observation by analyzing a light curve extracted from a source--free 
region, we used for the spectral analysis the background files accumulated
from long blank field exposures and available from the SDC public ftp site
(see Fiore et al. \cite{FGG99}, Parmar et al. \cite{Petal98}).

The PDS was operated in the customary collimator rocking mode, where half
collimator points at the source and half at the background and they are
switched every 96\,s. The PDS data were analysed using the XAS software
(Chiappetti \& Dal Fiume 1997) and the data reduction was performed according
to the procedure described in Chiappetti et al. (1999), inclusive of spike
filtering.
In the case of May data the attitude control software using a single
gyro was in a non favorable condition, resulting in significant gaps
in the reconstructed attitude data up to 20 min in each orbit while the
satellite pointed at the source (these intervals are rejected by the
standard event file creation software). However since the MECS is kept
on during such intervals, one can use XAS to accumulate images from
telemetry and verify that, despite a little blur, the satellite is
not drifting significantly. 
Considered also the sizeable flat top response of the PDS collimator,
it is therefore justified to use also such intervals in PDS spectra
accumulation, e.g. using a plain limit on Earth elevation angle above
5 degrees (this explains why the PDS exposure times in Table 1 are longer).
No such problem was present in June data (attitude gaps were small and
confined during Earth eclipses).

The source was not detected by the HPGSPC detector, thus we will not
discuss these data.

\subsection{Spectral Analysis}

For the spectral analysis, the LECS data have been considered only in the 
range 0.1--4 keV, due to still unsolved calibration problems at higher 
energies (Fiore et al. \cite{FGG99}). 
To fit the LECS, MECS and PDS spectra together, one has to introduce a 
constant rescaling factor to account for uncertainties in the 
inter--calibration of the instruments. 
The acceptable values for these constants are in the range 0.7--1.0 for the 
LECS and in the range 0.77--0.95 for the PDS, with respect to the MECS 
(Fiore et al. \cite{FGG99}). 
The spectral analysis was performed with the XSPEC 10.0 package.
As expected, during the May observation \object{ON\,231} was in a high state
with respect to previous X--ray observations (Comastri et al. 1997) and 
the source was detected also with the PDS up to 100 keV. 

\begin{figure}
\begin{center}
\begin{tabular}{c}
{\resizebox{\hsize}{!}{\includegraphics{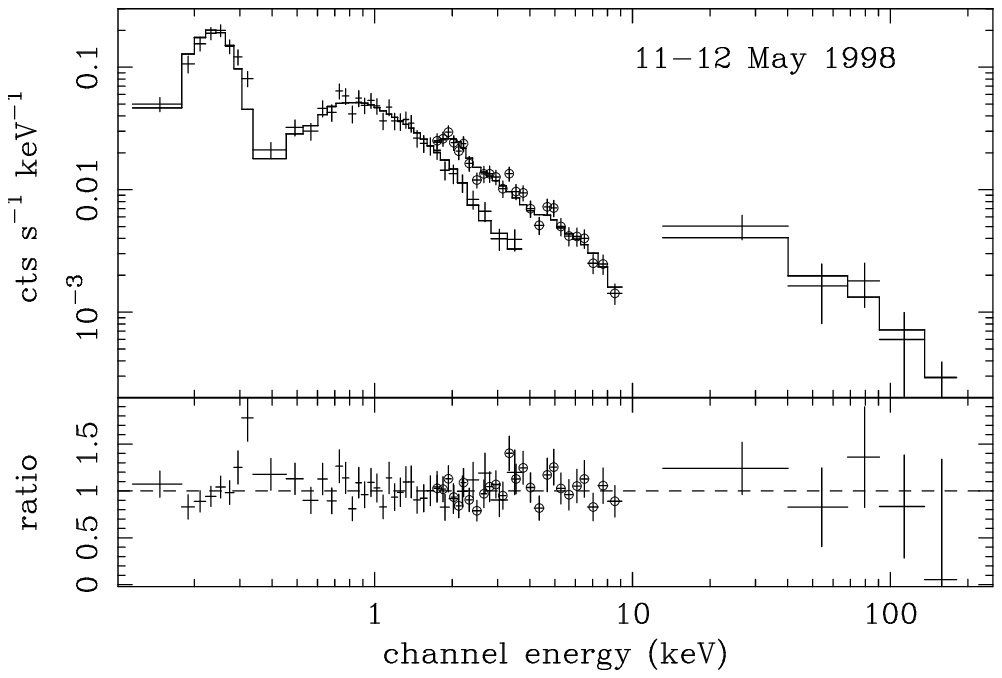}}} \\
{\resizebox{\hsize}{!}{\includegraphics{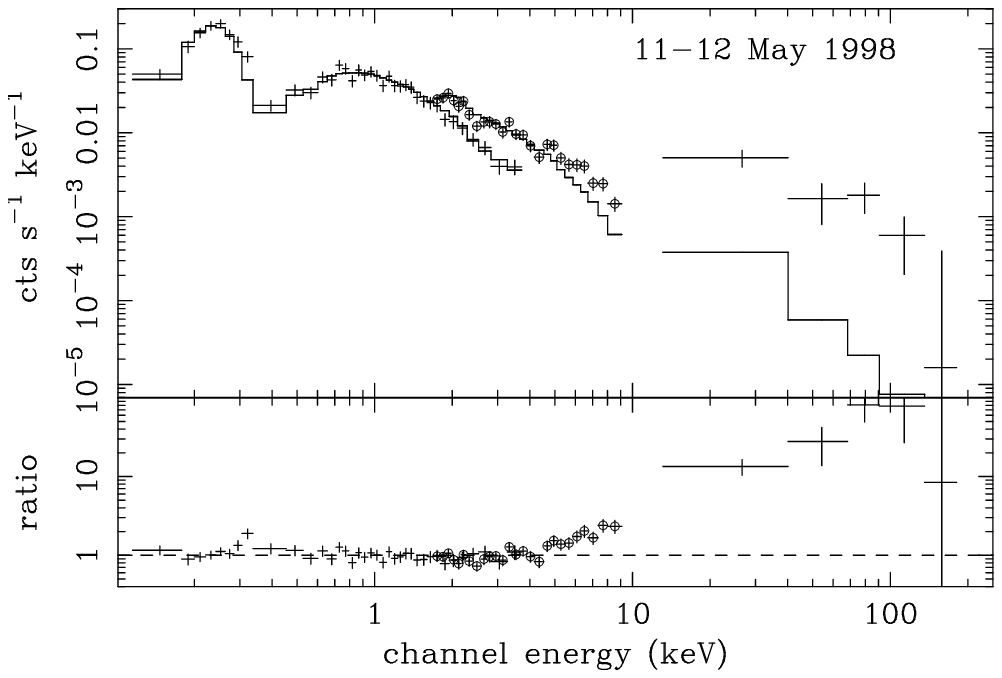}}} \\
\end{tabular}
\end{center}
\caption{Top panel: LECS+MECS+PDS \object{ON\,231} spectrum during the May
1998 observation, the best fit model is a broken power law.
Bottom panel: the same spectrum fitted with a power law model
(over the entire energy range), clearly showing that a simple
power law model can not fit the data.}
\label{fig:sp}
\end{figure}

In the fitting procedure we first considered only the LECS and MECS data. 
We fitted a single power law model plus absorption with the column 
density fixed at the Galactic value $N_{\rm H}=2\times 10^{20}$ cm$^{-2}$.
This model does not give a good fit to the data yielding a 
reduced $\chi^2_r=1.9$ (60 degree of freedom).
The fit did not improve by letting $N_{\rm H}$ free to vary.
In particular, the data at energies above $\sim 4$ keV could not be 
fitted by the same power law fitting the lower energies data
(see residual in Fig.\,\ref{fig:sp}, lower panel).
Instead, a broken power law (BPL) model provides a good fit to the data
with $\chi^2_r=0.96$ (58 degree of freedom), the best fit values and
error at 90\% (for three parameters of interest) are given in Table 2.
Notice that the spectrum hardens significantly at higher energies.

We then considered also the PDS data which approximately lie on the 
extrapolation of the BPL that fits the LECS \& MECS data.
By including also the PDS data in the fit procedure, the second photon spectral
index ($\Gamma_2$) is somewhat flatter than before, the break is at 
slightly higher energy and the errors on the second spectral
index are smaller (see Table 2). The flux at 1 keV is $1.7 \mu$Jy.
In Fig.\,\ref{fig:sp}, top panel, we report the LECS-MECS-PDS spectra
together with the BPL best fit. In this fit the LECS intercalibration constant
factor has an acceptable value of 0.80, while for the PDS we kept the
constant fixed at 0.9. In the same figure, lower panel, we report the best--fit
with a single power law for comparison.

\begin{table*}
\caption{Fit results for a broken power law and a two--power law model}
\begin{center}
\begin{tabular}{|ll|cclcccc|}
\hline
& & broken power law model \\
date & instruments & $\Gamma_1^{\rm \ a}$  & $\Gamma_2$ & break & 
$\chi ^2_r$ (d.o.f.) &$F_{[0.1-2~{\rm keV}]}^{\rm c}$ &$F_{[2-10~{\rm keV}]}$ 
&$F_{[10-100~{\rm keV}]}$ \\
& &  &  &  energy$^{\rm \ b}$ & & erg cm$^{-2}$ s$^{-1}$ & 
erg cm$^{-2}$ s$^{-1}$ & erg cm$^{-2}$ s$^{-1}$ \\
\hline
& & & & & & & & \\
11-12/5/98 &LECS-MECS     &$2.60^{-0.08}_{+0.08}$ &$1.27^{-0.63}_{+0.70}$ 
                          &$3.80^{-0.45}_{+0.80}$ &0.96 (58)       
                          &                   & &                        \\
           &LECS-MECS-PDS &$2.60^{-0.08}_{+0.07}$ &$1.13^{-0.20}_{+0.30}$ 
                          &$4.00^{-0.70}_{+0.55}$ &0.92 (63) 
                          &$2.3\times 10^{-11}$ & $4.4\times 10^{-12}$
                          &$3.2\times 10^{-11}$  \\ 
11-12/6/98 &LECS-MECS      &$2.68^{-0.12}_{+0.11}$ &$1.51^{-0.28}_{+0.27}$ 
                          &$2.55^{-0.60}_{+0.50}$ &0.88 (45) 
                          &                   & &                        \\
& LECS-MECS-PDS & $2.68^{-0.12}_{+0.11}$  & $1.49^{-0.26}_{+0.25}$ &
                $2.60^{-0.60}_{+0.50}$ & 0.88 (47) 
                & $1.7 \times 10^{-11}$ & $3.2 \times 10^{-12}$ 
                & $1.2 \times 10^{-11}$ \\
& & & & & & & & \\
\hline \hline
& & & & & & & & \\
& & two-power law model \\
& & $\Gamma_1$  & $\Gamma_2$ & n1/n2$^{\rm \ d}$ & $\chi ^2_r$ (d.o.f.) 
&$F_{[0.1-2~{\rm keV}]}$ &$F_{[2-10~{\rm keV}]}$ &$F_{[10-100~{\rm keV}]}$ \\
& &  &  & & & erg cm$^{-2}$ s$^{-1}$ & erg cm$^{-2}$ s$^{-1}$
& erg cm$^{-2}$ s$^{-1}$ \\
\hline
& & & & & & & & \\
11-12/5/98 &LECS-MECS-PDS     &$2.69^{-0.10}_{+0.09}$ &$0.84^{-0.12}_{+0.16}$ 
                          &$20^{-13}_{+22}$ & 0.96 (63) 
                          &  $2.2\times 10^{-11}$ 
                          & $4.2\times 10^{-12}$ &  $3.3 \times 10^{-11}$ \\

& & & & & & & & \\

11-12/6/98 &LECS-MECS-PDS     &$2.87^{-0.16}_{+0.14}$ &$1.11^{-0.30}_{+0.34}$ 
                          &$5.6^{-4.8}_{+7.5}$ &0.90 (47) 
                          & $1.6\times 10^{-11}$ 
                          & $3.2\times 10^{-12}$ &  $1.9 \times 10^{-11}$  \\
& & & & & & & & \\
\hline
\end{tabular}
\end{center}
$^{\rm a}$ photon spectral index; $^{\rm \ b}$ energy values in keV;
$^{\rm c}$ corrected for the absorption;
$^{\rm \ d}$ ratio between the two power law normalizations.
\end{table*}

During the June observation the source was detected in a state fainter than in
May, although still higher than earlier X--ray observations.
Again a BPL model was still necessary to fit  the LECS--MECS data (a fit 
with a simple power law gave a $\chi^2_r = 3.3$ for 47 dof). 
The first spectral index ($\Gamma_1$) is very similar to the previous 
observation, while the break is at lower energies. This suggests  that
the break moves toward lower energies when the source is weaker.
The second spectral index ($\Gamma_2$) seems steeper, but at 90\% confidence
level it is consistent with the value found for the May observation
(due to the poorer statistics at high energies, the second spectral index
$\Gamma_2$ is not very well constrained). The LECS constant is 0.85.
The PDS detection in June is less significant and it does not add significant
information. Again we kept the PDS rescaling constant factor fixed to 0.9
in the best-fit procedure. The inclusion of these data does not change the
results (see Table 2). The flux at 1 keV is $1.2 \mu$Jy.

\begin{figure}
\begin{center}
\begin{tabular}{c}
{\resizebox{\hsize}{!}{\includegraphics{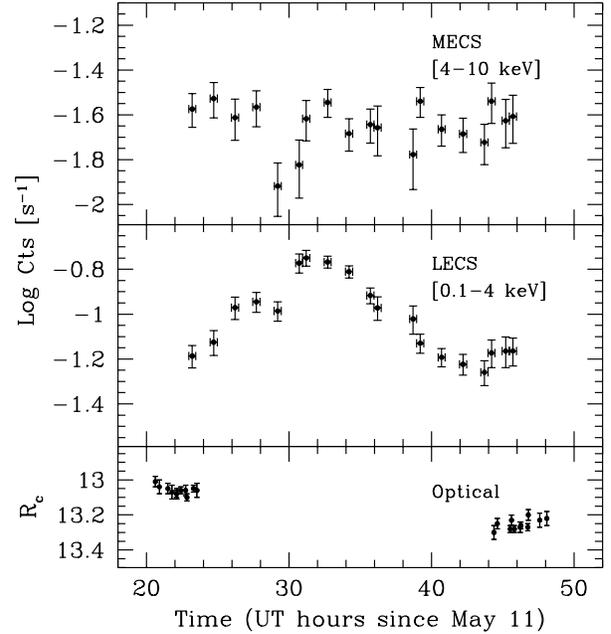}}}
\end{tabular}
\end{center}
\caption{MECS 4--10 keV (top panel) and LECS 0.1--4 keV (mid panel)
light curves of ON 231 during the May 1998 {\it Beppo}SAX observation.
The two figures have the same scale, note the different amount of variability.
The bottom panel shows the optical data in the $R_c$ band.
}
\label{fig:lc}
\end{figure}

Given the concave shape of our two spectra, we then fitted the sum of two power
law models, which is more physical than a concave broken power law. The results
are also reported in Table 2, together with the ratio between the two
power law normalizations. Formally the fit is as good as the BPL ones,
with the first spectral index steeper and the second one flatter.
The two power laws cross at about the values of the breaks found with
the BPL model.

\subsection{Time Variability}

As apparent from the best-fit fluxes reported in Table\,2, during the
second observation the source was weaker. We checked if the amount of
variability was different at energies lower or higher than the break. 
We considered the LECS and MECS detectors, which have a much higher
statistics than the PDS, and calculated the count rates in four different
energy bands: 0.1--2.0, 0.1--4.0, 2.0--10 and 5.0--10 keV, for the
two observations. The values for the May observation are: 
$0.077 \pm 0.002$, $0.086 \pm 0.002$, $0.050 \pm 0.002$, $0.014 \pm 0.001$.
For the June observation we have: 
$0.050 \pm 0.003$, $0.059 \pm 0.002$, $0.040 \pm 0.001$, $0.013 \pm 0.001$.
Thus, on a monthly time scale, the amount of variability in the energy
band 0.1-10 keV seems to be greater at softer energies.

In the May observation rapid X--ray variability of about a factor of three 
in 4--5 hours was clearly detected, but only at energies smaller than
3--4 keV, 
confirming a higher amount of variability at energies below the break. 
This can be seen from the light curves of Fig.\,\ref{fig:lc}: in the 0.1--4
keV band the flux increased by a factor $\sim$3 just after the starting 
of the observation and reached the maximum level at about 30h. 
This level was maintained for about 2--3 hours and then the count rate 
declined to $\sim$ 0.06 cts\,s$^{-1}$, comparable to the level measured at 
the beginning of the observation. Above 4 keV this variability, if at all
present, is much less pronounced. Note that our $3\, \sigma$ limit, in the
band 4--10 keV, corresponds to a variability of 40\%. 
Thus, at high energies, the source is much less variable than below 4 keV.

We extracted LECS and MECS spectra during the flare
(from the third to the ninth points of the X--ray light curve shown in Fig.\,2)
and outside the flare and performed the spectral analysis. 
Again, in both cases a power law did not fit the data and a BPL model was 
necessary. 
The first spectral index is steeper during the flare 
($\Gamma_1 = 2.7 \pm 0.06$ vs $2.4 \pm 0.15$ outside the flare). 
The break seems to move
at higher energies (best fit values are 4.4 and 3.5 keV, respectively), 
although the two values are consistent inside the 90\% confidence errors 
for three parameters of interest. 
The second spectral index does not change at all. 
Thus, also the fast variability that we detected 
during the May observation, suggests that the break moves at higher energies
when the source flux increases.

In the June observation we did not detect significant variability,
neither at high nor at low energies.

\section{Optical Observations}

\begin{figure}
\begin{center}
\begin{tabular}{c}
{\resizebox{\hsize}{!}{\includegraphics{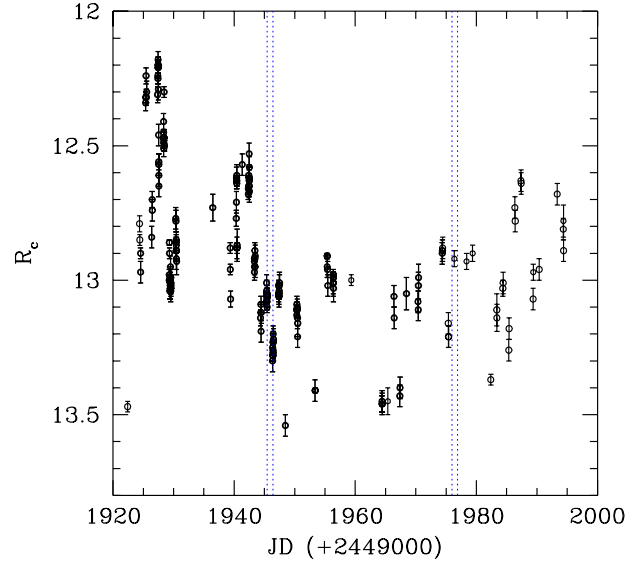}}}
\end{tabular}
\end{center}
\vskip -1truecm
\caption{Optical light curve for the period May--June, 1998. Data have
been obtained at the Perugia, Torino and Rome optical telescopes (see text).
The times of the two {\it Beppo}SAX observations are marked.}
\label{fig:optical}
\end{figure}

Optical photometry of ON 231 during the {\it Beppo}SAX pointings was
performed with some telescopes in Italy, in the standard bandpasses 
Johnson B, V 
and Cousins R, I, operated by the Perugia and Torino Observatories 
and by the Istituto Astronomico of University ``La Sapienza" in Roma. 
The main results of the optical observations during the great 1998 
outburst of ON 231 were already presented by Massaro et al. (1999).
A detailed description of the instrumentation and data reduction,
together with a complete data list up to 1998 June 9 can be retrieved
from the article by Tosti et al. (1999). 
The mean V, R$_{\rm c}$ and I$_{\rm c}$ magnitude are given in Table 1.
From these data we also evaluated the optical (energy) spectral 
index (assuming $A_V = 0.19$) 
which was found equal to $1.24 \pm 0.08$ for all observations.
In Fig.\,\ref{fig:optical} we show the optical light curve of ON 231
in the R band from the end of April to about the end of June with the
data obtained at the three observing sites. The times of the two X--ray
observations are marked.

From this light curve we can see that the R magnitude in the period
between the two $Beppo$SAX pointings was in the interval 13.0 -- 13.5: the
source remained quite bright but at a mean level fainter than that of 
great burst of the end of April. In May and June the angular distance
of ON 231 from the Sun was small and we were able to perform our optical 
observations only for few hours: in particular, the observations of May
were only at the beginning and at the end of the $Beppo$SAX pointing and then 
missed the ``flare'' observed in the soft X--rays (see Fig.\,2).
In June, the weather conditions allowed to observe ON 231 only at the
beginning of the $Beppo$SAX pointing.

\section{Discussion}

\subsection{SED}

\begin{figure*}{}
\begin{center}
\begin{tabular}{c}
{\resizebox{\hsize}{!}{\includegraphics{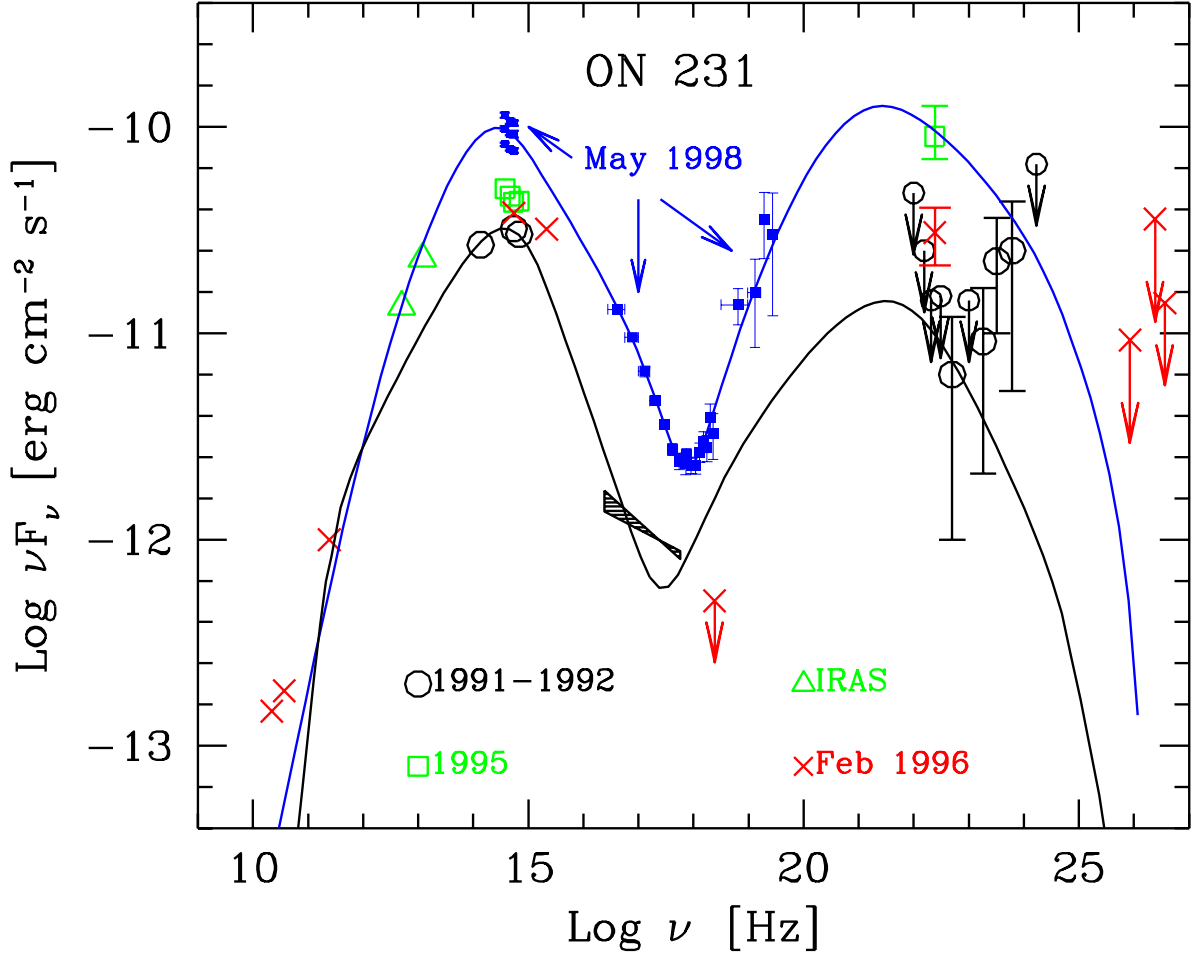}}}\\
\end{tabular}
\end{center}
\vskip -3.5 true cm
\caption{
The SED of ON 231, together with the 4 SSC models used to fit the far--IR 
to hard X--ray data. Source of data:
1991--1992: infrared and optical data (Feb. 1992) from Massaro et al. (1994);
  X--data (June 14, 1991) from Comastri et al. (1997), the spectral index
  in the ROSAT 0.1--2.4 keV band is $1.2\pm 0.05$;
  $\gamma$--ray data from Fig. 5 of von Montigny et al. (1995).
1995: optical data from Tosti et al. (1998);
  $\gamma$--ray data from Mukherjee et al. (1997).
1998: optical and X--ray data from this paper.
The lines are SSC models, whose input parameters are listed in Table 3.
}
\label{fig:sed}
\end{figure*}

In Fig.\,\ref{fig:sed} we show the spectral energy distribution (SED)
of ON 231, including our simultaneous X--ray and optical data.
The SED clearly shows that in the X--ray band we have detected,
simultaneously and with the same instruments, both the synchrotron
and the inverse Compton emission in the spectrum of a blazar.
Simultaneous detection of both components in the X-ray spectrum
of blazar have already been reported by Kubo et al. (1998) and by Giommi 
et al. (1999) for S5 0716+714, although not as clearly as in ON\,231.

In the same figure we plot other three sets of quasi--simultaneous
observations:
i) the data of the 1996 multiwavelength campaign as reported by Maisack
et al. (1997);
ii) the optical and $\gamma$--ray observations during 1995, when the source 
reached the brightest state in the EGRET band;
iii) the infrared, optical, X--ray and $\gamma$--ray data during 1991--1992, 
when the source was first detected in the $\gamma$--ray band by EGRET.
The last two sets of data are not strictly simultaneous (the $\gamma$--ray
fluxes detected in 1991--1992 refer to the sum of various pointings), but 
can illustrate the different states of the source.
The source was also detected by IRAS (Impey \& Neugebauer 1988), 
and the corresponding IR fluxes are reported in Fig.\,\ref{fig:sed},
even if they are not simultaneous with any other observations.

Note that there are some inconsistencies between the data in 1991--1992
as reported in Table 1 of von Montigny et al. (1995) and the fluxes reported 
in Fig. 5 of the same paper, which are consistent with the flux 
reported by Sreekumar et al. (1996).
We have reported the data as shown in Fig. 5 of von Montigny et al. (1995).
As can be seen, the 1991--1992 $\gamma$--ray spectrum is extremely hard
($\alpha\sim 0.4\pm 0.4$).
We could not find the spectral index for the 1995 $\gamma$--ray flux,
but the shape of the spectrum combining all observations together 
(from 1991 to 1995) is $\alpha\sim 0.73 \pm 0.18$ (Hartman et al. 1999), 
suggesting that the combined spectrum is steeper than the 1991--1992 spectrum
(and suggesting that the 1995 spectrum is steeper still).
Also shown are the upper limits in the TeV band, as derived by WHIPPLE
and HEGRA observations (Maisack et al. 1997) during Jan--Feb 1996.

Alike other blazars, the SED of ON 231 is characterized by two broad 
components, the first peaking at IR--optical frequencies and the second 
in the $\gamma$--ray band.
The first is believed to be synchrotron emission by a relativistic jet,
while the second component has been interpreted as synchrotron self--Compton
scattering,
possibly including some contribution from seed photons produced externally
to the jet (see e.g. Ghisellini \& Madau 1996 and reference therein) or 
synchrotron by ultra--relativistic electron--positron pairs generated by 
relativistic protons (the proton blazar model, Mannheim 1993, see Maisack 
et al. 1997 for the application of this model to ON 231).

The only other spectral information in the X--ray band is from ROSAT:
Comastri et al. (1997), using a single power law model, found an energy 
spectral index $\alpha_x=1.2\pm 0.05$, in agreement with Lamer et al. (1996).
An earlier determination of the spectral shape using Einstein data
resulted in an unconstrained spectral index (Worrall \& Wilkes 1990).
The shape of the X--ray spectrum at the time of the ROSAT observation
seems different from the one determined by $Beppo$SAX.
This could be due both to the narrower spectral coverage of ROSAT
and to the fact that the source was in a weaker state.
The 0.1--2.5 keV ROSAT spectrum, which is flatter than that 
measured by us in the same energy band (see index $\Gamma_1$ in Table 2),
could be due to the contribution, in this
energy band, of the flat component that $Beppo$SAX sees above 4 keV
in May and above 2.5 keV in June. If the break moves at lower energy
when the source is weaker, as suggested by the two $Beppo$SAX observations,
then during the ROSAT observation, when the source was more than a factor of 4
weaker, the break should be inside the ROSAT band, or even at lower energies.
In this case the spectrum detected by ROSAT would be either a combination 
of synchrotron and Compton emission or purely due to Compton scattering,
explaining the relative flatness of the ROSAT spectrum. A steepening of
the spectrum going to softer energies in the SED is also required by
the quasi simultaneous IR--optical data (Massaro et al. 1994).

\begin{table*}[ht]
\begin{center}
\caption{Input parameters for the SSC models shown in Fig.\,\ref{fig:sed}.}
\begin{tabular}{|r l|cccccccc|}
\hline
\hline
Date & &$R$  &$B$  &$L^\prime_{\rm inj}$  &$\delta$ 
&$\gamma_{\rm max}$   &$\gamma_{\rm min}$   &$s$  &$\beta_{\rm esc}$ \\
     & &cm   &G    &erg s$^{-1}$      &         
&    &            &                 &              \\
\hline
&&&&&&&&& \\
1998 &May 11  
&8$\times 10^{15}$ &0.86 &2.2$\times10^{42}$ &14.2 
&$1.1\times 10^5$ &$2.3\times 10^3$ & 2.9 &0.38 \\ 
1991 & &8$\times 10^{15}$ &0.83 &3.3$\times10^{41}$ &14.2 
&$7.0\times 10^4$ &$3.3\times 10^3$ & 3.5 &0.01 \\
\hline
\hline
\end{tabular}
\end{center}
\end{table*}

\subsection{Limits on magnetic field and particle energies}

The observed ``flare'' in the soft X--ray band is symmetrical 
(equal rise and decay timescale), suggesting that the variability 
timescale is determined by the light crossing time of the emitting 
region, $R/c$ (see Chiaberge \& Ghisellini 1999).
This in turn implies that the cooling time is shorter than $R/c$,
allowing us to put limits on the value of the magnetic field and on the
energy of the electrons producing the variable flux at the oberved
frequency $\nu_x$.
Using $t_{\rm var}=5$ hours and $\nu_x=3\times 10^{16}$ Hz, we derive 
$B>0.4\delta^{-1/3}$ Gauss and $\gamma_x< 1.5\times 10^5\delta^{-1/3}$.
Here $\delta=[\Gamma-\sqrt{\Gamma^2-1}\cos\theta]^{-1}$ 
is the Doppler beaming factor, where $\theta$ is the viewing angle.
Since the peak of the synchrotron emission must be at a frequency
$\nu_{\rm s} < 3\times 10^{14}$ Hz, the corresponding Lorentz factor of the 
electrons emitting at the peak is 
$\gamma_{\rm s}< 1.5\times 10^4\delta^{-1/3}$.
This implies that the peak of the self Compton flux is at 
$h\nu_{\rm c} = (4/3)\gamma^2_{\rm s}h\nu_{\rm s} < 370 \delta^{-2/3}$ MeV.

Tighter constraints can be obtained assuming that the optical and X--ray 
emission are cospatial, and that also the cooling time of the optical emitting 
electrons is shorter than the light crossing time.
Massaro et al. (1999) report intranight variations in the I and B bands
of 0.2 magnitude in 1 hour, with an approximately symmetric time profile.
Assuming again $t_{\rm var}=5$ hours, we obtain
$B>1.5\delta^{-1/3}$ Gauss and $\gamma_{\rm s}< 9.7\times 10^3\delta^{-1/3}$.

\subsection{Homogeneous SSC model}

The far IR to $\gamma$--ray emission from blazars can be explained by 
simple homogeneous and one--zone synchrotron inverse Compton models, 
with the emitting region moving relativistically towards the observer.
The inverse Compton emission may have two components: the first is
produced by relativistic electrons scattering off locally produced
synchrotron photons (SSC), while the second corresponds to the scattering
of photons produced in other regions (EC), either elsewhere in the jet
or by the broad emission line clouds, or by some scattering plasma
within these clouds.
Ghisellini et al. (1998), analyzing all blazars of known redshift
detected by EGRET with spectral information in the $\gamma$--ray band,
found that the EC component decreases its contribution as the total 
luminosity decreases, with lowest luminosity BL Lac objects requiring a 
negligible amount of EC.
We applied a pure SSC model to the SED of ON 231,
as shown in Fig.\,\ref{fig:sed}.
In particular we have tried to explain the different 
SED by changing the minimum number of parameters.

The applied models assume to continuously inject, in a spherical source 
of radius $R$ embedded in a tangled magnetic field $B$, relativistic 
electrons with a power law energy distribution $\propto \gamma^{-s}$, 
between $\gamma_{\rm min}$ and $\gamma_{\rm max}$.
The total luminosity injected in the form of relativistic electrons
is $L^\prime_{\rm inj}$, calculated in the comoving frame.
We also assumed to observe the source at the viewing angle $1/\Gamma$,
so that the Doppler factor $\delta=\Gamma$.
The steady--state particle distribution is the result of the injection 
and cooling processes, and we also account for possible escape of the 
particles, which may be relevant for ON 231.
It is assumed that particles escape at some velocity 
$v_{\rm esc}=c\beta_{\rm esc}$, independent of their energy.
Further details about this model can be found in Ghisellini et al. (1998).
The input parameters for the models shown in Fig.\,\ref{fig:sed} are
given in Table 3.
The size of the source and the Doppler factor have been kept fixed; 
the slope $s$ of the injected electron distribution
and $\gamma_{\rm max}$ are similar.
The magnetic field does not change,
while $\gamma_{min}$ changes by $\sim 30\%$, from 2300 to 3000.
The largest change concerns the injected power, increasing from 
$L^\prime_{\rm inj}= 3.3\times 10^{41}$ erg s$^{-1}$ (1991 model), to 
$L^\prime_{\rm inj}= 2.2\times 10^{42}$ erg s$^{-1}$ (1998 model), an increase
of a factor 7.
Within the SSC model, it is not possible to account for 
the very hard $\gamma$--ray spectrum of 1991--1992.
The quasi--simultaneous optical data requires the peak of the optical 
emission to be at frequencies between the near IR and the optical, 
while the hard EGRET spectrum indicates that the Compton peak is
at energies greater than a few GeV.
This translates in a lower limit for the energy of the electron emitting at
the peaks of $\gamma_{\rm peak}>6\times 10^4$.
These electrons emit at $\nu_{\rm s,peak}\sim 2\times 10^{14}$ Hz
if the magnetic field $B<1.4\times 10^{-2}\delta^{-1}$ Gauss.
This is not consistent with the limits derived in the previous section.
In addition, a small magnetic field would imply a very large radiation to 
magnetic energy density ratio, $U_{\rm r}/U_{\rm B}$, and then
an excessive self Compton flux, unless the Doppler factor is exceedingly 
large ($\delta> 100$, see eq. 2.5 and 2.6 in Ghisellini et al. 1996).
We then conclude that either the hard 1991--1992 
$\gamma$--ray spectrum is due to another
source, or, if it will be  confirmed to be associated with ON 231,
is produced by another component (i.e. to inverse Compton
scattering with photons produced externally to the jet).

Note that BL Lacertae showed a similar behavior (hardening of
the $\gamma$--ray spectrum) during the flare of summer 1997 
(Bloom et al. 1997). This was interpreted (Sambruna et al. 1999;
Madejski et al. 1999; Bottcher \& Bloom 1999) 
as due to an increased contribution of emission line
photons to the inverse Compton scattering process.
Something similar could have happened also to ON 231 (during the
1991--1992 EGRET observation), but the lack of spectroscopic
observations preclude any further conclusions.

What is interesting, and peculiar, in ON 231 is the sharp flattening, 
above $2-4$ keV, of the X--ray spectrum.
A population of electrons which cools only radiatively can not account for 
spectra as flat as observed in ON 231: the flattest predicted spectrum 
in the case of radiative cooling has an energy spectral index 
$\alpha=0.5$ (see e.g. Ghisellini et al. 1998).
We therefore must invoke an additional mechanism.
One likely possibility is escape.
In this case high energy electrons would cool before escaping, 
while low energy electrons would preferentially
escape before cooling radiatively.
The corresponding steady--state particle distribution 
would then show a flattening towards the low energy part, accounting 
for the very flat inverse Compton component emerging above 4 keV.
The model we have applied to ON 231 indicates that 
$\beta_{\rm esc}\sim 0.3$--$0.4$, corresponding to an escape time of 
the order of 2--3 light crossing times $R/c$.

The variability predicted by the model can account for the observed 
variability in the soft X--ray band and for the much less variable 
hard X--ray flux, even if the bolometric luminosity does not change.
This can be achieved by changing (even by a small amount) the slope $s$
of the injected electron distribution, without changing the total injected
power. This will change the synchrotron spectrum above the synchrotron
peak (characterized by $\alpha\sim s/2$), but not the 
flux below, nor the self Compton flux below the Compton peak, produced
by low energy electrons scattering low frequency synchrotron photons.

The $Beppo$SAX data of ON 231 show that this source
can be considered a BL Lac object intermediate between the HBL and
LBL sources.
The concave shape of the X--ray spectrum in the 0.1--10 keV
band can be used to define the class of intermediate blazars,
which should be characterized, in this band,
by the presence of the steep tail of the synchrotron radiation
and the hard emerging of the Compton emission.

\acknowledgements{This research was financially supported by the Italian
Space Agency. The Roma and Torino groups acknowledges financial support
from the Italian Ministry for University and Research (MURST) under the
grants Cofin98-02-03 and Cofin98-02-32. We thank the {\it Beppo}SAX
Science Data Center (SDC) for their support in the data analysis.}

\end{document}